\documentstyle[12pt]{article}

\newcommand{\xf}{{\bf x}^f}
\newcommand{\xin}{{\bf x}^{in}}

\newcommand{\beq}{\begin{equation}}
\newcommand{\eeq}{\end{equation}}
\newcommand{\beqn}{\begin{eqnarray}}
\newcommand{\eeqn}{\end{eqnarray}}
\newcommand{\eq}[1]{(\ref{#1})}
\newcommand\Appendix[1]{\par
\setcounter{section}{0}
 \setcounter{equation}{0}
 \renewcommand{\thesection}{Appendix \Alph{section}}
\section{#1}
 \def\theequation{\Alph{section}.\arabic{equation}}}

\title{
\vspace{-2.0cm}
\begin{flushright}
\begin{large}
ITEP-TH-11/95\\
quant-ph/9511009
\end{large}
\end{flushright}
\vspace{1.5cm}
Pauli-Potential and Green Function Monte-Carlo Method
for Many-Fermion Systems}

\author{B.L.G.~Bakker,\\
{\it Department of Physics and Astronomy, Vrije Universiteit}\\
{\it De Boelelaan 1081, NL-1081 HV Amsterdam, The Netherlands} \\
        M.I.~Polikarpov and A.I.~Veselov \\
{\it ITEP, B. Cheremushkinskaya 25, Moscow, 117259 Russia}
       }

\begin{document}
\sloppy
\date{}
\maketitle
\begin{abstract}
The time evolution of a many-fermion system can be described by a Green's
function corresponding to an effective potential, which takes
anti-symmetrization of the wave function into account, called the
Pauli-potential.
We show that this idea can be combined with the Green's Function Monte Carlo
method to accurately simulate a system of many non-relativistic fermions.
The method is illustrated by the example of systems of several (2-9)
fermions in a square well.

\end{abstract}

\section{Introduction}

The application of Green's Function Monte-Carlo (GFMC) algorithms for the
simulation of bosonic and fermionic systems is well known \cite{Kalos},
\cite{Ceperley}.  However, the fermionic case is much more difficult to deal
with than the bosonic one \cite{Ceperley}, \cite{Scalapino}, \cite{Koonin}.
In the framework of simulations of many-fermion systems employing the
Langevin equation, Tursunov and Zhirov \cite{TuZh89} introduced the idea of
a Pauli-potential, to account for the repulsive forces between fermions due
to anti-symmetrization. We study the implementation of this idea in the more
efficient GFMC method.

In this work we first describe the standard GFMC method, then we discuss the
proposal of Tursunov and Zhirov. After that we illustrate the implementation
of the proposed  approach within the frame work of GFMC calculations of the
simple quantum mechanical system of several fermions in a square well. With
moderate computational effort we simulate nine spinless fermions

\section{The Green's Function Monte Carlo Algorithm}

In the paper \cite{TuZh89} the authors applied the Langevin-equation to
study multi-fermion systems. We use a more efficient method: Green's
Function Monte Carlo \cite{Kalos}, \cite{Ceperley} in a modified form
\cite{KePoSh90}. The idea of the GFMC-method is to determine the
ground-state energy and wave function by operating iteratively with the
Green's function on an arbitrary function. The Green's function itself is
obtained as a solution of the standard resolvent integral equation:
\begin{equation} G(E) = G_T (E) + G(E) \, (V-V_T ) \, G_T (E) .  \label{eq5}
\end{equation} The (exactly known) Green's function $G_T(E)$ is the
resolvent for the Hamiltonian with the potential $V_T$. The time evolution
of the system is determined by the propagator in the time representation.
For imaginary time this propagator is the Laplace-transform of $G(E)$.
Because we solve eq.~(\ref{eq5}) by iteration using the standard MC method,
in order to achieve fast convergence, it is important to employ a trial
potential $V_T$ that is as close as possible to $V$.  Besides this trial
potential, there are other elements in the GFMC-method that make that method
so efficient compared  to other stochastic methods. These features are: the
guidance function $\Psi_G$ that guides the MC process and the trial energy
$E_T$; the details of the standard GFMC-method are described in the
Appendix.

	We use the modified algorithm proposed in ref.\cite{KePoSh90}; which
allows to work with the integral equation of the type \eq{eq5} even in
the case where the kernel is not positive definite.
In the standard approach the wave-function
is represented by a set of points, which move randomly and may disappear or
reappear with some multiplicity, proportional to the kernel of the integral
equation (see step 8 of the algorithm described in the Appendix).
In the modified GFMC method the
multiplicity $m_I$ (see \eq{mI}) is proportional to the absolute value
of the kernel, and all points which are going to the intermediate generation
with that multiplicity, also change their phase: $\delta(x) \rightarrow
\delta(x) + \delta_K$, where $\delta_K$ is the phase of the kernel $K$ which
enters into the definition of $m_I$ \eq{mI}. To explain this
modification we give a simple example. To calculate the ``expectation
value'' $<\phi> = \int \phi(x)f(x) \, dx/\int f(x) \, dx $
for the complex function $f(x)$ we can generate
the set of points $\{x_k\}$ with the probability proportional to $|f(x)|$,
and calculate $<\phi> =
\sum_k \phi(x_k) e^{i \delta_k}/\sum_k e^{i \delta_k}$ where $\delta_k$
is the phase of $f(x_k)$. Of course, the convergence of this
procedure is not guaranteed for all choices of $\phi$ and $f$.
Our results show that for the case studied here (fermions in a
square well) there exists a sufficiently broad range of parameters of the
algorithm, in which convergence is obtained.

The simple improvement described above allows for inclusion of the sign of
the wave function: the density of points corresponds to the magnitude of
the wave function. The phase of a point corresponds to the phase of the
wave function at that particular position. In this way we are also able to
accommodate dynamical nodes in the wave function, which occur in fermionic
systems already in the ground state.
Our algorithm provides the dynamical nodes in the wave function
owing to the action of the Pauli-potential $V^F$ \eq{eq4}.

\section{The Idea of Tursunov and Zhirov}

The main complication for the use of stochastic simulations of multi-fermion
systems is the fact that its wave function must be anti-symmetrized. A basic
tool of the methods used here is the imaginary-time single-particle
propagator

\beq
 U\left( {\bf x}^f , {\bf x}^{in} ; \beta \right) =
 C \exp \left
 \{ - \frac{m \left( \xf - \xin \right)^2}{2  \beta } -
  \beta V \left(\frac{\xf + \xin}{2}\right) \right\},
 \label{eq1}
\eeq
where $m$ is the mass of the particle and $\beta$ is a (small) time step.
For two identical fermions, 1,2, the imaginary-time propagator can be
written as:

\begin{eqnarray}
 U^{(F)} \left( \xf_1 , \xf_2 , \xin_1 , \xin_2 ; \beta \right) & = &
 \nonumber \\
  U^{(D)} \left( \xf_1 , \xf_2 , \xin_1 , \xin_2 ; \beta \right)& - &
  U^{(D)} \left( \xf_2 , \xf_1 , \xin_1 , \xin_2 ; \beta \right) ,
 \label{UU}
\end{eqnarray}
where (F) refers to fermions and (D) to distinguishable particles.
$U^{(D)}$ is the product of two single-particle
propagators:

\beq
U^{(D)} \left( \xf_1 , \xf_2 , \xin_1 , \xin_2 ; \beta \right) =
 U\left( {\bf x}^f_1 , {\bf x}^{in}_1 ; \beta \right)
U\left( {\bf x}^f_2 , {\bf x}^{in}_2 ; \beta \right) \label{Uf}
\eeq
At small values of $\beta$,
anti-symmetrization (\ref{UU}) can be effectively implemented \cite{TuZh89}
by using an additional effective potential:

\beqn
U^F(1,2) & = & U^D(1,2) - U^D(2,1) = U^D(1,2) \left(1 - U^D(2,1)/U^D(1,2)
\right) \nonumber \\
 & \approx & U^D(1,2) e^{- \beta V^F(1,2)}.
\eeqn
This additional
potential, $V^{(F)}$, has the following form to leading order in
$\beta$:
\beq
 V^{(F)}(1,2) =
  - \frac{1}{\beta} \ln \left[ 1 - \exp \left( - \frac
 {m (\xf_1 - \xf_2 ) \cdot (\xin_1 - \xin_2 )}{\beta} \right) \right] .
 \label{eq3}
\eeq
We see that the Pauli-exclusion principle leads to a complex, nonlocal and
time-dependent potential; still, it can be used in the Monte-Carlo algorithm.

For the N-fermion case we will have
\begin{eqnarray}
 V^{(F)}(1, \ldots , N)  = &
 \label{eq4} \\
  & \frac{-1}{\beta} \ln \left[ 1 - \exp \left( -
 \sum_{k<l} \frac {m (x^f_k - x^f_l ) \cdot (x^{in}_k - x^{in}_l )}{\beta}
 \right) \right] . \nonumber
\end{eqnarray}
So, to leading order in $\beta$, this ``Pauli-potential'' corresponds to
anti-symmetrization of pairs of particles only: the sum in eq.~(\ref{eq4})
has only N(N-1)/2 terms.  It is very important that the permutations of
three and more particles occur when the time-development of the system is
simulated by repeated operation of the propagator $U^{(F)}$, {\it i.e.}, by
the repeated action of the potential $V^F$ \cite{TuZh89} during the Monte
Carlo procedure.

\section{Results}

We obtained our results by solving eq.~(\ref{eq5}) in the
time-representation.  Using this Green's function, we obtain the
ground-state wave function of the system dependent on time: $\psi_0 (x;
\beta) \sim \exp ( -E_0 \beta) \psi_0 (x; 0)$. For an N-particle system in
three dimensions, the wave function is represented in the MC-method by a set
of points in 3N--dimensional space. The density of the points in such a
population follows modulus of the wave function $| \psi_0 (x; \beta) |$. The
action of the density matrix (Laplace-transformed Green's function) is
performed in finite imaginary time steps $\beta$.

In order to obtain the Green's function, we perform a Laplace transform of
the density matrix by sampling $\beta$ from the distribution
$\exp(-\beta / \Delta)/\Delta$.  We varied $\Delta$ in order to be able to
extrapolate our results to the point $\Delta = 0$, which means that,
on average, the time step $\beta$ tends to zero, since we used the
Pauli-potential only to leading order in $\beta$.

In order to determine the energy $E_0$ of the ground state, we monitor the
size of the population in time and use \eq{Egp} from the Appendix.

We found that we could decrease the fluctuations in the energy by an order
of magnitude if we would kill all points that have a multiplicity (see items
4 and 5 of the GFMC algorithm described in the Appendix)
larger than some value $M_{max}$ (Typically $M_{max} \approx 5 - 20$).
By this procedure we introduce a systematic error proportional to
$\Delta$. Doing so, we do not change the character of the systematic error:
it remains linear in $\Delta$.


For a check of our approach we have, until now,
studied the case of several spinless fermions in a square-well potential.
The number of particles in this well was varied between two and nine.

For the trial potential $V_T$ we use the oscillator interaction, for which
the Green's function is known in closed form \eq{Rhot} \cite{FEYNMAN}.  For
the guidance function $\psi_G$ we use a Slater determinant of
harmonic-oscillator wave functions.  (These may or may not correspond to the
same oscillator as is used for the trial potential.)

In Fig.1 we show for a value of $\Delta = 0.0005$ the development of the
energy for a system of nine fermions with the number of time steps. In this
calculation we use a square well with depth $V_0 = -3.5$, radius $R = 2$;
the number of points in each population (which describes the nine fermion
wave function) was approximately one thousand.  Clearly, the energy
converges to a value of -12.8, which differs from the true value (-11.501),
indicated by the broken line.  The reason for this phenomenon is the fact
that we still have a finite $\Delta$.

In order to see the effect of
taking smaller time steps, we calculated the energy for different values
of $\Delta$.
Fig.2 shows this dependence of the average energy for the same problem
on the size of $\Delta$. We clearly see that the average energy tends to
the exact energy if $\Delta$ tends to zero.
Extrapolation of the energy values to the point $\Delta = 0$ gives a value
that is almost equal, within the error bars to
the exact value for this case (-11.501). These simulations were performed for
a value of $E_T = -10.5$, differing from the exact value of the energy, to
mimic a realistic situation, when the value of the exact energy is not known.

The true value of the energy can be found by extrapolating the computed
values to $\Delta=0$.  We illustrate this for the case of nine bodies in
Fig. 3.  The number of killed points (taking into account their
multiplicities) divided by the total number of points in our simulation
process is plotted as a function of $\Delta$. We see that this number indeed
depends linearly on $\Delta$.  A value $M_{max}=5$ was taken.  We checked
that a similar linear dependence occurs when we change  $M_{max}$.

The dependence of the average energy on $\Delta$ is shown in Fig. 4
for the case of five fermions in the same square-well potential.
The linear dependence of the energy on $\Delta$ is clearly seen to occur for
sufficiently small values of $\Delta$.

We also found that the Pauli potential acts in such a way that the motion of
points
that render the argument of the logarithm in eqs.~(\ref{eq3},\ref{eq4})
negative, is hindered.  We checked, by recording the
number of points that do cross the border, that the Pauli repulsion
effectively blocks the crossing, for $\Delta\beta \rightarrow 0$.

\section*{Acknowledgement}

The authors are grateful to L.V. Shevchenko for numerous fruitful
discussions; and to O.O. Tursunov and O.V. Zhirov for stimulating remarks
when this work was started.  MIP and AIV wish to express their gratitude to
the members of the Department of Physics and Astronomy of the Free
University at Amsterdam, where most of the work described here was done.
They were partially supported by the Netherlands Organization for Scientific
Research, the International Science Foundation, grant No. MJM300, and by the
Russian Foundation for Fundamental Sciences (Grant No. 93-02-03609).

\Appendix{ }

Below we describe the main steps of the algorithm \cite{Ceperley}
(see also the pedagogical paper \cite{LEE} or the book by Kalos  and
Whitlock \cite{WHITLOCK} )
for the solution of
the integral equation \eq{eq5}.
\begin{enumerate}

\item A set of points is sampled from a distribution $f_1(x) = |\Psi_G(x)|^2$.
For $N$ fermions in $D$ dimensions $x$ is a $D N$-dimensional vector
representing the positions of $N$ particles (if we exclude the
center of mass motion it is an $(N-1)D$-dimensional vector).
For the initial distribution
we take $|\Psi_G(x)|^2$, because eventually we will obtain from the process
the distribution $\Psi(x) \Psi_G(x)$ rather than $\Psi(x)$. The typical
number of points in this set, called the initial generation, is several
hundreds.

\item It is convenient to work with the density matrix
$\rho(x,x',\beta) = \sum_n\Psi_n^*(x)\Psi_n(x')e^{- E_n \beta}$. The density
matrix
is related to the Green's
function $G(x,x',E)$ by the Laplace transform. In our algorithm we
carried out the Laplace transform  by sampling the imaginary time $\beta$
from the distribution $\frac{1}{\Delta} \exp\left\{\ -\frac {\beta}{ \Delta}
\right\}$.  
found that the additional random number occurring in the sampling of the
imaginary-time distribution involved in the Laplace transform, improves the
statistical errors.

\item  To each point $x'$ in the initial generation the diffusion and
drift is applied, after which the points are distributed with the
probability

\beq
f(x,\beta) = \int \rho_D(x,x',\beta) f_1(x') \,dx',
\label{f}
\eeq
where
\beq
\rho_D(x,x',\beta) = \left(\frac {m} {2\pi\beta} \right)^{\frac{D N}{2}}
\exp\left\{ - \frac{m(x-x'-\beta F(x'))^2}{2\beta}\right\}
\eeq
m is the mass of the fermion and the ``quantum force'' $F$ is given by:

\beq
F_i(x) = \frac1m \Psi_G^{-1}(x)\frac{\partial}{\partial x_i} \Psi_G(x),
\eeq
Note that $F_i(x)$ is the component of a vector as
$\frac{\partial}{\partial x_i}$ is.

Because $\rho_D$ does not enter the final expression, as
the dependence on $\rho_D$ is cancelled due to the fact
that $\rho_D$ enters also
in the denominator of $m_D$ (see \eq{mD}) the random walk of points can be
performed with an arbitrary probability distribution.
For the distance probability distribution a Gaussian form is used,
since random numbers with a Gaussian
distribution can be generated very efficiently on a computer. The second
reason for choosing this distribution is, that it is similar
to the real distribution corresponding to the bound state.

Algorithmically the sampling of the distribution \eq{f} is done in two steps.
First we shift the initial
points: $x''= x' + \beta F(x')$, and after that we add to each point the
gaussian random numbers $\eta_i$ with unit expectation value:
$x_i = x''{_i}+ \sqrt{\frac{2\beta}{m}}\eta_i $

\item In order to construct the new generation, multiple copies of
each point $x$ are produced. The multiplicity is given by the formula:

\beq
m_D(x,x',\beta) = e^{E_T \beta}\frac{\Psi_G(x)}{\Psi_G(x')}
\frac{\rho_T(x,x',\beta)}{\rho_D(x,x',\beta)} \label{mD}
\eeq

where $\rho_T$ is the density matrix
(the Laplace-transformed Green's function) for the trial potential
which satisfies the well known equation:

\beq
 \frac{ \partial \rho_T (x, x', \beta)}{\partial \beta} =
 - H_T \rho (x, x', \beta);
 \; \rho_T(x, x', 0) = \delta(x - x').
\eeq

For the trial hamiltonian we use the harmonic oscillator hamiltonian, for
which $\rho_T$ is known explicitly \cite{FEYNMAN}:

\beqn
 \rho_T(x, x', \beta)  & =  &
\left( \frac{m \omega_T} {2 \pi \sinh \omega_T \beta}
 \right) ^{\frac{N D}{2}}  \label {Rhot} \\
 & \times & \exp \left\{ - \frac{m \omega_T}
 {2 \sinh \omega_T \beta} [(x^2 + x'^2) \cosh \omega_T \beta - 2 x \cdot x']
 -c_T \beta \right\} \nonumber
\eeqn

This trial density matrix corresponds to a hamiltonian with the potential
$V_T$, given by

\beq
V_T (x) = \frac{m \omega_T^2 x^2}{2} + c_T.
\eeq

The trial energy $E_T$ is introduced in order to avoid exponential growth
or shrinkage of the number of points of the population.
As a result, the number of points in the population fluctuates around
a value that has the time-dependence $\exp\{ -(E_0 - E_T) \beta \}$.
Effectively, this amounts to a shift in the hamiltonian with the constant
energy $E_T$.

\item The other, so called intermediate, branch of the process is
formed by creating another set of multiple copies of the points $x$:
for them the multiplicity is given by:
\beq
m_I(x,x',\beta) = \frac{K(x,x',\beta) m_D(x,x',\beta) \Delta}
{\rho_T(x,x',\beta)}, \label{mI}
\eeq

here $K$ is the Laplace transformed kernel of the integral equation
\eq{eq5},

\beq
K(x,x',\beta) = [V_T(x)-V(x)] \rho_T(x,x', \beta).
\eeq

In general $m_D$ and $m_I$ are not integers. We convert them to
integers by adding a uniformly distributed random number to each of them
and take the integer part.

\item Each intermediate point, created this way, is treated in the same way
as the points taken from the initial generation, {\it i.e.}, they will take
part in the random walk with branching until they are eventually propagated
to the new generation. If the average value of $m_I$ is less than unity, this
process is completed in a finite time on the computer.

\end{enumerate}

The sequence of operations described above, correspond to the terms in the
iterative solution of the equation:

\beq
 \tilde{\rho} = \tilde{\rho}_T + \Delta \cdot K \ast \tilde{\rho}
 \label{eq6}
\eeq
where $K = V$ and $\ast$ denotes the
integration over the intermediate coordinates:
$K \ast \tilde{\rho} = \int K(x,x'') \tilde{\rho}(x'',x') \, dx''$.
The direct points correspond to the term $\tilde{\rho}_T$, while the
intermediate points correspond to $\Delta \cdot K$.  If an intermediate
point is processed again, it may be promoted immediately to the new
generation, in which case it corresponds to the term $\Delta \cdot
\tilde{\rho}_T \ast K$, otherwise it will be an intermediate point again,
now corresponding to $\Delta^2 \cdot K \ast K$ etc. The difference in the
normalization of the kernels of the equations \eq{eq5} and \eq{eq6} is due
to the difference in the normalization of $G$ and $\tilde{\rho}$:

\beqn
\tilde{\rho}(x,x') & = & \frac{1}{\Delta}
 \frac{\Psi_G(x)}{\Psi_G(x')}
G(x,x',E_T-\frac{1}{\Delta}) \nonumber \\
 & = & \frac{\Psi_G(x)}{\Psi_G(x')}
\sum_n\frac{\Psi_n^*(x)\Psi_n(x')}{1+\Delta(E_n-E_T)} \label{rhot}
\eeqn
here $\Psi_n(x)$ and $E_n$ are the wave
function and the energy of the n-th
level of the Hamiltonian studied here.
The distribution
of points in the new (second) generation is sampled with the probability
distribution

\beq
f_2(x) = \int \tilde{\rho}(x,x')  f_1(x') \, dx'.
\eeq
In the same way we can get the distributions $f_3,...,f_n$;

Using expression  \eq{rhot} for $\tilde{\rho}(x,x')$
it is easy to prove that as $n\rightarrow \infty$
$f_n(x) \rightarrow const.\Psi_G (x)\Psi_0(x)/
[1 + \Delta (E_0 -E_T)]^{n-1} + ...$,
where $\Psi_0(x)$ and $E_0$ are the  wave function and the energy
of the ground state. So the ground state energy can be calculated
from the number of points, $P_n$ in the $n$-th generation:

\beq
	E_0 = E_T + \frac{1}{\Delta}\left(\frac{P_{n-1}}{P_n} - 1\right).
 \label{Egp}
\eeq

\newpage

\section*{ Figure Captions}

\begin{itemize}

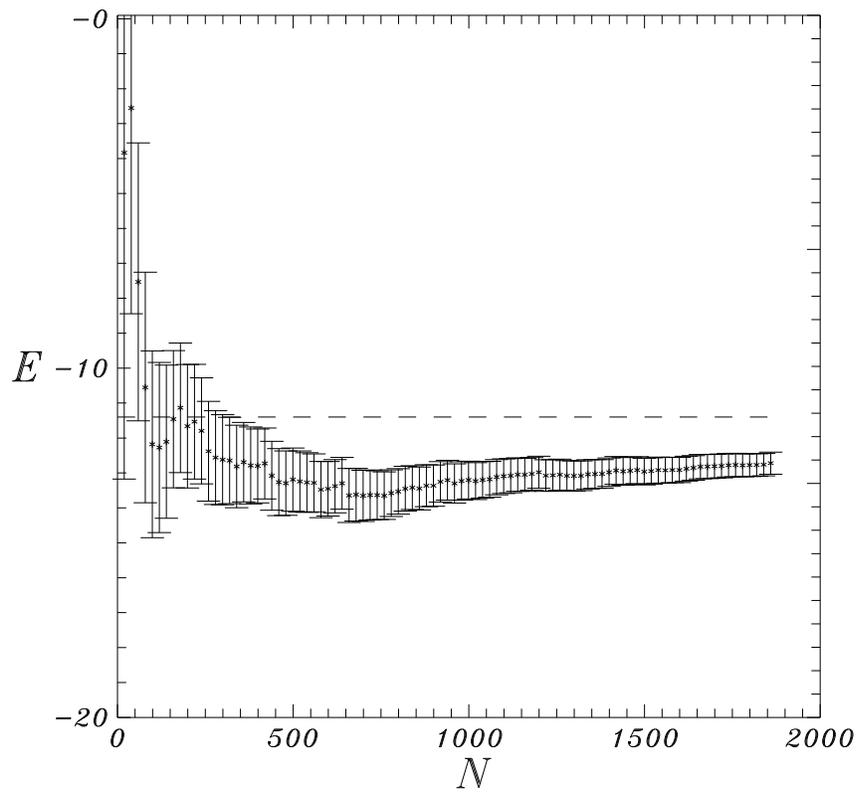
\item[Fig. 1] The development with the number of time steps ($N$) of the
energy for a system of nine fermions. The value of $\Delta$ is 0.0005.

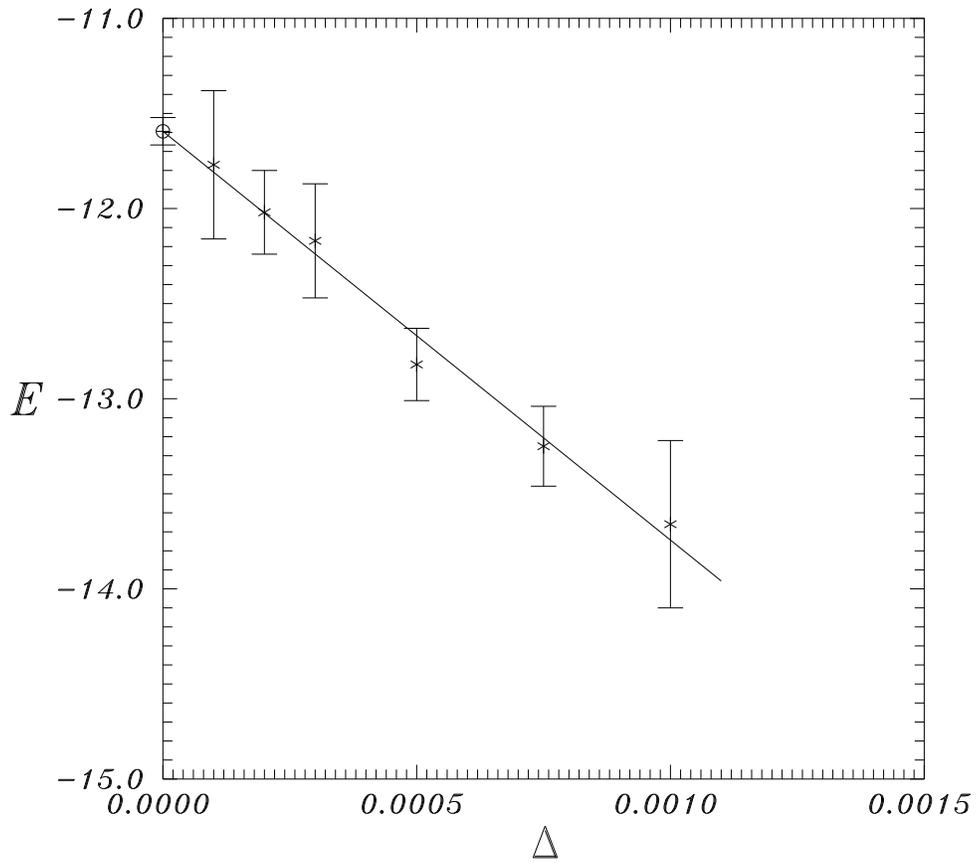
\item[Fig. 2] Dependence of the energy of the system of nine fermions on the
time step $\Delta$. A linear fit is made, which gives an estimate of the
energy for $\Delta = 0$.

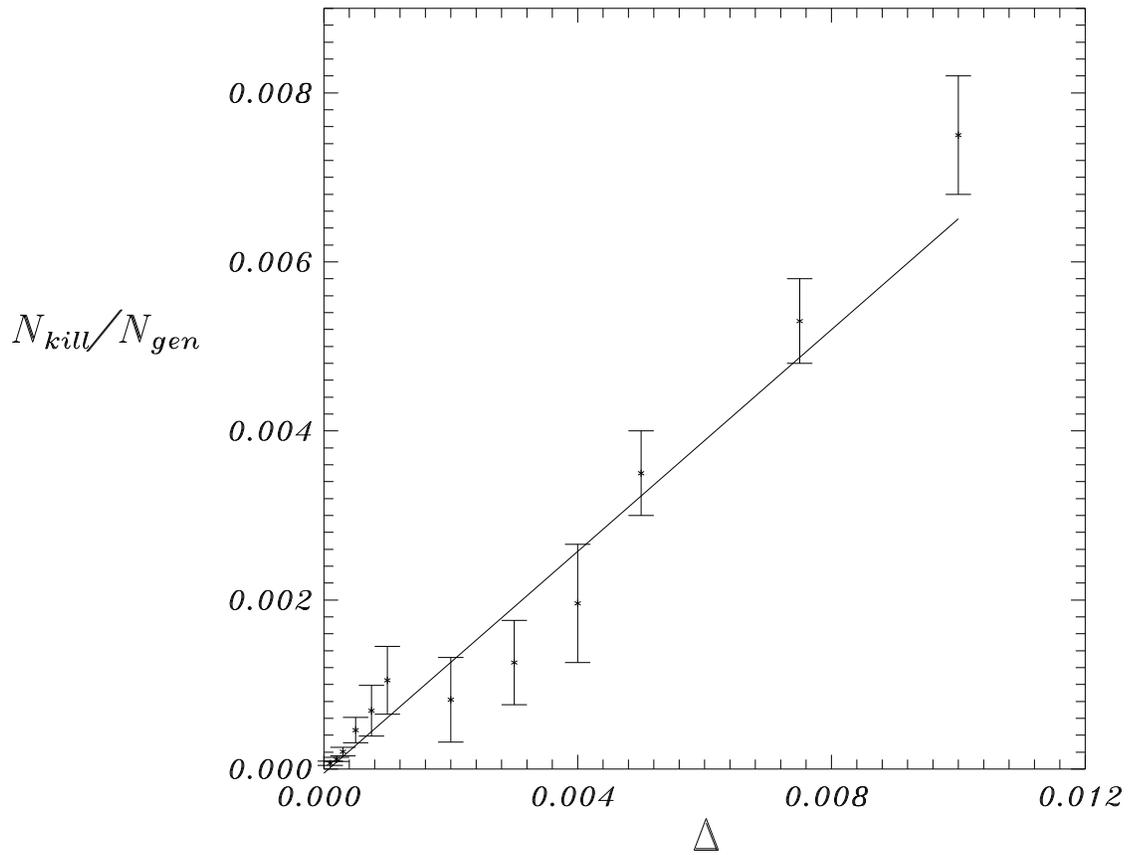
\item[Fig. 3] The dependence on $\Delta$ of the number of killed points for
the system of nine fermions divided by the total number of points in our
simulation. The parameter $M_{max} = 5$.

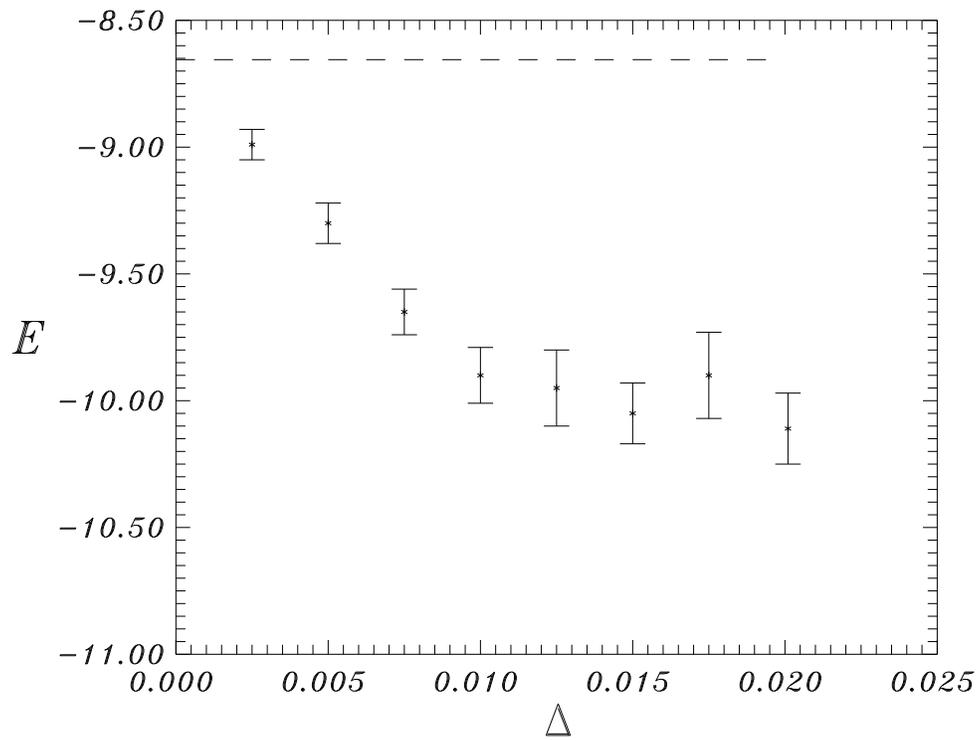
\item[Fig. 4] The same as Fig. 2, but now for five fermions. It is seen that
the linear dependence of the energy on $\Delta$ obtains for small values of
$\Delta$. A linear fit of the points at low values of $\Delta$
intersects the axis at the exact energy (indicated by the broken line).

\end{itemize}

\end{document}